# Comb-based WDM transmission at 10 Tbit/s using a DC-driven quantum-dash mode-locked laser diode

Pablo Marin-Palomo,[1,*] Juned N. Kemal,[1] Philipp Trocha,[1] Stefan Wolf,[1] Kamel Merghem,[2] François Lelarge,[3] Abderrahim Ramdane,[2] Wolfgang Freude,[1] Sebastian Randel[1] and Christian Koos[1,4,**]

[1]*Institute of Photonics and Quantum Electronics (IPQ), Karlsruhe Institute of Technology (KIT), 76131 Karlsruhe, Germany*
[2]*Centre de Nanosciences et de Nanotechnologies (CNRS), Univ. Paris-Sud, Université Paris-Saclay, C2N – Avenue de la Vauve, 91220 Palaiseau cedex, France*
[3]*Almae Technologies, Route de Nozay, 91140 Marcoussis, France*
[3]*Institute of Microstructure Technology (IMT), Karlsruhe Institute of Technology (KIT), 76344 Eggenstein-Leopoldshafen, Germany*
*\*pablo.marin@kit.edu*, *\*\*christian.koos@kit.edu*

**Chip-scale frequency comb generators have the potential to become key building blocks of compact wavelength-division multiplexing (WDM) transceivers in future metropolitan or campus-area networks. Among the various comb generator concepts, quantum-dash (QD) mode-locked laser diodes (MLLD) stand out as a particularly promising option, combining small footprint with simple operation by a DC current and offering flat broadband comb spectra. However, the data transmission performance achieved with QD-MLLD was so far limited by strong phase noise of the individual comb tones, restricting experiments to rather simple modulation formats such as quadrature phase shift keying (QPSK) or requiring hard-ware-based compensation schemes. Here we demonstrate that these limitations can be over-come by digital symbol-wise phase tracking algorithms, avoiding any hardware-based phase-noise compensation. We demonstrate 16QAM dual-polarization WDM transmission on 38 channels at an aggregate net data rate of 10.68 Tbit/s over 75 km of standard single-mode fiber. To the best of our knowledge, this corresponds to the highest data rate achieved through a DC-driven chip-scale comb generator without any hardware-based phase-noise reduction schemes.**

## 1. Introduction

With the explosive growth of data rates across all network levels [1], wavelength-division multiplexing (WDM) schemes are becoming increasingly important also for short transmission links that connect, e.g., data centers across metropolitan or campus-area networks. In this context, optical frequency combs have emerged as particularly attractive light sources, providing a multitude of narrowband optical carriers at precisely defined frequencies that may replace tens of even hundreds of actively stabilized laser diodes [2-8]. Among the various comb generator concepts, chip-scale devices are of special interest, in particular when it comes to short-distance WDM links, for which compactness and cost-efficient scalability to high channel counts is of utmost importance.

Over the previous years, a variety of chip-scale comb generators have been shown to enable high-speed WDM transmission at Tbit/s data rates. The highest performance was achieved by exploiting Kerr nonlinearities in integrated optical waveguides, either for spectral broadening of initially narrowband frequency combs [7] or for Kerr comb generation in high-Q microresonators [4-6]. While these approaches allow to generate hundreds of spectral lines distributed over bandwidths of 10 THz or more, the underlying setups are rather complex, require high pump power levels along with delicate operation procedures, and still comprise discrete components such as fiber amplifiers or external light sources. Other approaches to chip-scale comb generators rely on gain switching of injection-locked DFB lasers [9-11], which may be integrated into chip-scale packages with all peripheral components but suffer from the rather small bandwidth of the overall comb, which typically spans less than 500 GHz [11].

These limitations may be overcome by so-called quantum-dash (QD) mode-locked laser-diodes (MLLD) [12-21]. These devices combine an inhomogeneously broadened gain spectrum of the QD material with passive mode-locking through four-wave mixing [12,13] and allow for generation of broadband flat frequency combs. QD-MLLD can be operated by a simple DC current while producing combs that typically comprise 50 carriers or more, distributed over a spectral range of more than 2 THz. However, the devices suffer from large optical linewidth of the individual tones, which strongly impairs transmission of signals that rely on advanced modulation formats [14-17]. As a consequence, data transmission demonstrations with QD-MLLD have so far either been limited to, e.g., simple quadrature phase-shift keying (QPSK) [14,15] or have required dedicated hardware schemes for reducing the optical linewidth of the comb tones [16-19].

In this work, we show that the phase-noise limitations of QD-MLLD-based WDM transmission schemes can be overcome by a continuous feed-forward algorithm for symbol-wise phase tracking [22], without any additional optical hardware. Expanding on our earlier demonstrations [15], our concept allows for data transmission using 16-state quadrature amplitude modulation (16QAM), even in the presence of strong phase noise of the QD-MLLD tones. The viability of our concept is demonstrated in a series of transmission experiments. In a first experiment, we transmit 52 channels over 75 km of standard single-mode fiber (SSMF) using QPSK as a modulation format. By ex-ploiting polarization-division-multiplexing (PDM), we achieve an aggregate line rate (net data rate) of 8.32 Tbit/s (7.83 Tbit/s). In a second experiment, we use 38 carriers to transmit PDM-16QAM signals over 75 km of SSMF at an aggregate line rate (net data rate) of 11.55 Tbit/s (10.68 Tbit/s). To the best of our knowledge, this is the highest data rate achieved by a DC-driven chip-scale comb generator that does not require any hardware-based phase-noise reduction schemes. Our investigation is supported by an in-depth analysis of the phase-noise characteristics of the QD-MLLD, revealing a strong increase of the FM-noise spectrum at low frequencies as the main problem that has prevented 16QAM transmission so far. Our results combined with the robustness and the ease of operating QD-MLLD show the great potential of such devices as light sources for highly scalable WDM transceivers.

## 2. Quantum-dash mode-locked laser diodes

The frequency comb sources used in our data transmission experiments consist of InAs/InGaAsP quantum-dash (QD) structures, which are grown by molecular beam epitaxy (MBE) on InP substrates [23]. The active region consists of 6 stacked layers of InAs QD (bandgap corresponding to a photon energy at a wavelength of $\lambda_g = 1.5 \ldots 1.6\,\mu m$), embedded into InGaAsP barriers ($\lambda_g = 1.17\,\mu m$), see Fig. 1(a). Details of the dash-in-a-barrier design can be found in [12]. The thickness of the barrier between the QD layers amounts to 40 nm, and the structure is terminated by 80-nm-thick separate confinement heterostructure (SCH) layers of InGaAsP ($\lambda_g = 1.17\,\mu m$) towards the top and bottom InP regions. Electrons are injected into the SCH layer from the $n$-side, corresponding to the bottom contact in Fig. 1(a), and are then trapped in the QD, emitting photons with a wavelength of approximately 1.55 µm. The photons are guided in a buried ridge waveguide of 1.5 µm width [12]. Cleaved facets act as broadband front and backside mirrors, thereby forming a Fabry-Perot (FP) laser cavity with an FSR of approximately 42 GHz, corresponding to a total length of 980 µm. Due to the inhomogeneously broadened gain originating from the shape distribution of the QD, multiple longitudinal modes can coexist in the laser cavity. According to [23], mode-locking in QD-MLLD can be attributed to mutual sideband injection due to self-induced carrier density modulations at the FSR frequency, similar to actively mode locked laser diodes, but induced by the signal generated from the beating among the laser cavity modes rather than by an external radio frequency (RF) source. This leads to a comb of equally spaced spectral lines with strongly correlated phases, thereby forming a pulsed output signal with a repetition rate corresponding to the cavity round-trip time. The strong phase correlation of neighboring modes leads to a narrow RF beat note measured at the FSR frequency when detecting the pulsed output with a high-speed photodetector

To identify favorable parameters for operating the QD-MLLD in transmission experiments, we performed a thorough characterization of the devices using the setup shown in Fig. 1(b). The light emitted by the DC-driven QD-MLLD is collected by a lensed single-mode fiber (LF), which features an anti-reflection coating to avoid any distortions by optical back-reflection into the laser cavity. The collected light is sent to an optical spectrum analyzer (OSA) and to a photodiode (PD), which is connected to an electrical spectrum analyzer (ESA) for extracting the RF beat note at the FSR frequency. In our experiment, we swept the pump current and recorded the fiber-coupled output power of the device along with the linewidth of the RF beat note and the optical spectrum, from which we extract the number of lines within a 3 dB bandwidth of the comb, see Fig. 1(c) and (d). Figure 1(d) shows the full-width half-maximum (FWHM) linewidth of the RF beat note at the FSR frequency, the fiber-coupled output power, and the number of comb lines within the 3 dB bandwidth of the comb, all as a function of the injection current, for a constant temperature of the QD-MLLD of approximately 21.2 °C. A favorable operation regime can be found between 300 and 420 mA (shaded in red), featuring low RF linewidth along with high optical power and more than 30 comb lines within the 3 dB-bandwidth of the spectrum. We repeated the measurement for other temperatures between 20 and 25 °C, observing a similar trend. Figure 1(c) shows the resulting frequency comb spectrum for an injection current of 390 mA at a stabilized temperature of 21.2 ºC. During operation, the injection current and temperature at which the laser is set are kept constant to avoid any drift of the center frequency of the generated combs.

For application in coherent optical communications, the carriers used to transmit data need to have high optical carrier-to-noise power ratios (OCNR) and narrow optical linewidths, or, equivalently, low phase noise. The OCNR values of the QD-MLLD tones can be extracted from the optical spectrum of the frequency comb, Fig. 1(c). For carriers within the 3 dB-bandwidth of the comb, we find OCNR values of at least 37 dB using the usual reference bandwidth of 12.5 GHz, which corresponds to a frequency span of 0.1 nm at a center wavelength of 1550 nm. The OCNR was determined from the spectrum shown in Fig. 1(c) by estimating the power spectral density of the noise background by taking into account the resolution bandwidth (RBW) of the spectrum analyzer (ANDO AQ6317, RBW = 12.5 GHz). The recorded OCNR level would in principle safely allow for transmission of 16QAM signals at symbol rates of 50 GBd or more [24]. However, the individual comb tones exhibit strong phase noise, which prevents the use of advanced modulation formats with high spectral efficiency. In the case of strong phase noise, there is a high probability that the induced phase change causes a wrong recovery of the symbol, thus dramatically increasing the bit-error ratio (BER), even at high OCNR. In the next section, we present a more detailed analysis of the phase noise for individual carriers of the QD-MLLD, whilst Section 4 contains an analysis of the impact of phase noise on the measured BER for QPSK and 16QAM modulation formats.

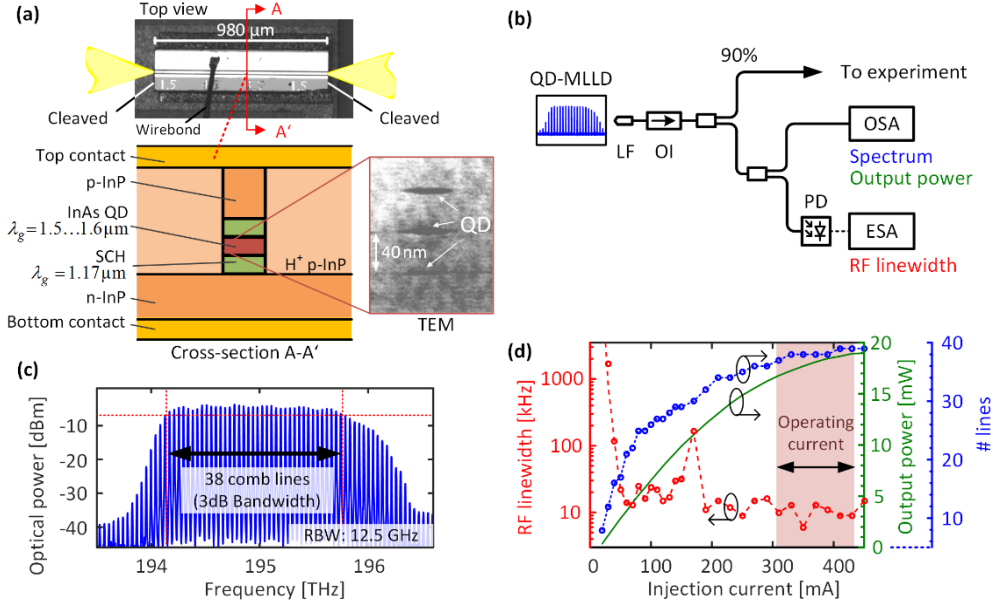

**Fig. 1**. Frequency-comb generation in quantum-dash mode-locked laser diodes (QD-MLLD). **(a)** Top view and cross-section schematic of a QD-MLLD consisting of a ridge waveguide of 1.5 µm width and 980 µm length. The active region comprises 6 stacked layers of InAs QD (bandgap corresponding to a photon energy of $\lambda_g$ = 1.5 ... 1.6 µm), separated by 40 nm-thick InGaAsP barriers ($\lambda_g$ = 1.17 µm), see transmission-electron microscope (TEM) image in the Inset [12]. Carriers are injected into the active region through 80-nm-thick separate confinement heterostructure (SCH) layers of InGaAsP ($\lambda_g$ = 1.17 µm). Proton implantation of the p-InP region ($H^+$ p-InP) ensures electrical isolation **(b)** Setup for frequency comb characterization. The QD-MLLD is driven by a DC current. LF: Lensed fiber. OI: Optical isolator. PD: Photodiode. ESA: Electrical spectrum analyzer. OSA: Optical spectrum analyzer. **(c)** QD-MLLD frequency comb spectrum for an injection current of 390 mA at a stabilized temperature of 21.2 °C. **(d)** Optical output power (green), number of lines within the 3-dB bandwidth of the comb (blue), and FWHM of the RF beat note at the FSR frequency (red) of 42 GHz as a function of the injection current. Shaded region: Operating current for low RF linewidth, high output power and large number of lines.

## 3. Phase noise characteristics of QD-MLLD

The optical tones of the QD-MLLD are broadened by several effects [25]. Fundamentally, laser linewidth broadening is caused by the coupling of spontaneous emission into the oscillating mode, leading to spectrally white phase noise. Other effects such as flicker and random-walk frequency noise contribute to linewidth broadening, too, and occur at longer time scales [26,27], i.e., lower frequencies. Influences of temperature fluctuations and mechanical vibrations happen on an even larger time scale and are disregarded. In the following, we differentiate between short-term and long-term phase noise, leading to a short-term and a long-term linewidth of the comb tone. Short-term phase fluctuations are characterized by a white frequency-noise spectrum and lead to a Lorentzian laser line shape with a FWHM $\Delta f_L$, while long-term phase fluctuations lead to a Gaussian spectrum with FWHM $\Delta f_G$ [26].

For measuring the short-term and long-term optical linewidths of the individual comb lines, we follow two different heterodyne approaches, which are illustrated in Fig. 2(a), Setup I and Setup II. In Setup I, the QD-MLLD output is superimposed with the output of a highly stable tunable local-oscillator laser (LO I) and then sent to a photodetector connected to an ESA for recording the beat-note spectrum and the long-term linewidth. In Setup II, we use a second tunable local oscillator laser (LO II) along with a coherent receiver and high-speed analog-to-digital converter (ADC). The ADC is used to record the high-frequency beat note, from which the intrinsic Lorentzian, i.e., short-term linewidths are extracted by offline processing of the signals [26], see Section 3.2. In both cases, the LO linewidths must be much smaller than the expected linewidth of the investigated comb line. This is verified by connecting LO I to Input II (Inp II) of Setup II, and LO II to Input I (Inp I) of Setup I and by observing that in both cases the resulting linewidths are much smaller than those obtained from the QD-MLLD tones. In the following sections, we detail a quantitative phase-noise model, Section 3.1, and use it to extract the QD-MLLD phase-noise characteristics from the measurements obtained from Setup I and Setup II, Sections 3.2 and 3.3.

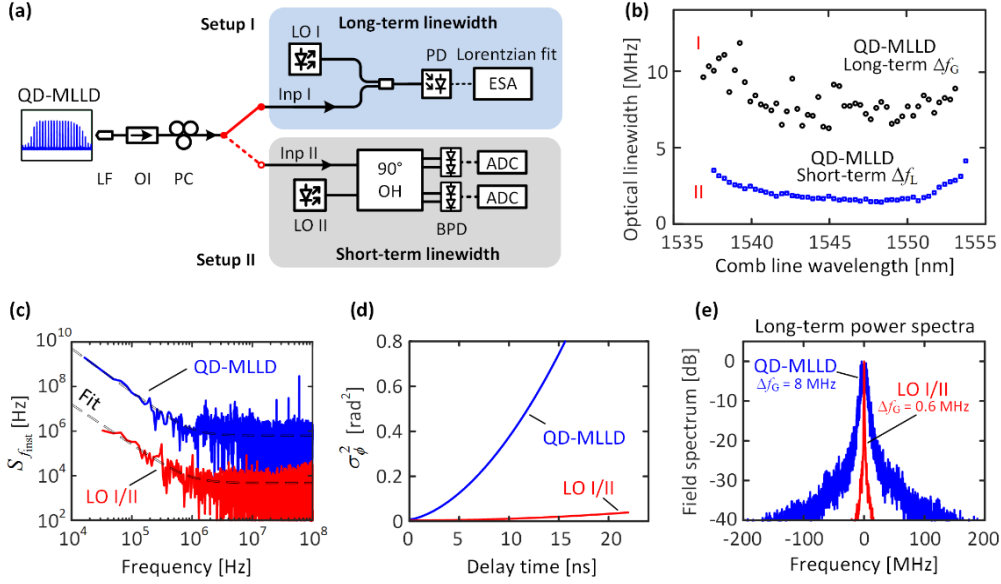

**Fig. 2.** Linewidth measurement and phase-noise characterization for selected comb lines of the QD-MLLD. **(a)** Setups for linewidth measurement and phase-noise characterization. In Setup I, the QD-MLLD output is superimposed with the output of a narrowband local oscillator laser (LO I) and detected by a single photodiode (PD) and an electrical spectrum analyzer (ESA) for measuring the long-term linewidth $\Delta f_G$. In Setup II, a second LO laser (LO II) is used along with a coherent receiver and high-speed ADC for measuring the short-term phase fluctuations and the associated intrinsic Lorentzian linewidth $\Delta f_L$. LF: Lensed fiber. OI: Optical isolator. PC: Polarization controller. LO I, LO II: Tunable external cavity lasers (ECL). OH: Optical hybrid. BPD: Balanced photodiodes. ADC: Analog-to-digital converter. Inp I, Inp II: Auxiliary inputs for verification of the LO laser linewidths. **(b)** Long-term and short-term optical linewidths $\Delta f_G$ and $\Delta f_L$ of the different comb lines. The long-term linewidths were recorded with an observation time of $\tau_0 \approx 15 \, \mu s$. **(c)** Power spectra $S_{f_{inst}}$ of the instantaneous frequency fluctuations (FM noise) for the QD-MLLD and an ECL tone along with model fit according to Eq. (1). The data was obtained by testing a QD-MLLD tone with Setup II (MLLD, blue) or by connecting LO I to Inp II of Setup II (LO I/II, red). In both cases, the wavelength of the tested tone was 1547.3 nm. Note that the second measurement can only reveal the relative phase fluctuations of LO II with respect to LO I, which can be considered as an upper boundary of the phase noise of each of the sources. Since the linewidth of LO I is much smaller than that of the QD-MLLD tone, the phase fluctuations in the first measurement can be attributed to the QD-MLLD tone. **(d)** Phase-noise variance $\sigma_\phi^2$ as a function of measurement time $\tau$, both for LO I and for the QD-MLLD tone. **(e)** Power spectra as a function of the frequency offset from the carrier for LO I and for the QD-MLLD tone (RBW = 200 kHz).

### 3.1. Phase-noise and linewidth model

The complete statistical characteristics of the phase noise or, equivalently, the frequency noise of a laser oscillator can be obtained from the so-called FM-noise spectrum $S_{f_{inst}}$, i.e., the power-spectral-density function of the instantaneous optical frequency fluctuations $f_{inst}(t)$, which can be modeled by [28,29]

$$S_{f_{inst}}(f) = S_L + S_1 f^{-1} + S_2 f^{-2}. \quad (1)$$

This power spectral density is composed of spectrally constant white frequency noise $S_L f^0$, flicker frequency noise $S_1 f^{-1}$, and random-walk frequency noise $S_2 f^{-2}$. In these relations, the quantities $S_L$, $S_1$ and $S_2$ are the constant coefficients of the various frequency noise contributions.

In the presence of flicker and random-walk frequency noise, the resulting laser line can be approximated by a Gaussian if the instantaneous optical frequency fluctuations $f_{inst}(t)$ are dominated by low-frequency components at frequencies $f$ that fulfill $S_{f_{inst}}(f) > (4f \ln 4)/\pi^2$ [30], i.e., for frequencies $f$ that are smaller than an upper frequency $f_{high}$. The spectral power beyond this point is small and does usually not influence the center part of the line shape since the noise level of $f_{inst}$ is small compared to the Fourier frequency $f$. Towards smaller Fourier frequencies, the lowest frequency $f_{low}$ to be measured is given by the observation time $\tau_0 = 1/f_{low}$. For $f_{high} > 5 f_{low}$, a good approximation of the long-term FWHM linewidth is then [30]

$$\Delta f_G = [8 \ln(2) A]^{1/2}, \quad (2)$$

where the parameter $A$ corresponds to the spectral power of the frequency-noise between $f_{low}$ and $f_{high}$,

$$A = \int_{f_{low}}^{f_{high}} S_{f_{inst}}(f) \, df, \quad (3)$$

In the presence of flicker or random-walk frequency noise, $A$ increases with decreasing lower frequency limit $f_{\text{low}}$, i.e., the long-term linewidth broadens with increasing measurement duration $\tau_0 = 1/f_{\text{low}}$. In contrast to that, the line shape is a Lorentzian if flicker and random-walk frequency noise are absent and the frequency noise spectrum is white, i.e., $S_1 = S_2 = 0$. In this case, the intrinsic Lorentzian linewidth $\Delta f_{\text{L}}$ is directly proportional to the constant spectral power density $S_{\text{L}}$ of the frequency noise [30],

$$\Delta f_{\text{L}} = \pi S_{\text{L}}, \tag{4}$$

Measuring the FM-noise spectrum and fitting Eq. (1) to the measurement allows us both to obtain the intrinsic linewidth $\Delta f_{\text{L}}$, see Section 3.2, and to determine the long-term linewidth $\Delta f_{\text{G}}$, see Section 3.3. The results are shown in Fig. 2(b).

### 3.2. Short-term linewidth from measured FM-noise spectrum

The intrinsic Lorentzian linewidth $\Delta f_{\text{L}}$ for each line of the frequency comb is obtained from the FM-noise spectrum $S_{f_{\text{inst}}}$ using Eqs. (1) and (4). To measure $S_{f_{\text{inst}}}$, we first recover the in-phase and quadrature component of the electric field of individual comb carriers with a coherent receiver [26] using Setup II of Fig. 2(a). The intermediate frequency of the down-converted complex electric field is selected by tuning LO II, and the phase differences $\phi(t)$ between the comb line and LO II are then evaluated in an intermediate frequency band between 1 GHz and 2 GHz. The time-dependent instantaneous frequency fluctuation $f_{\text{inst}}(t)$ are obtained from the time-derivative of the measured phases,

$$2\pi f_{\text{inst}}(t) = \frac{d\phi(t)}{dt} \approx \frac{\phi(t+\tau_s) - \phi(t)}{\tau_s}, \tag{5}$$

where $\tau_s$ denotes the sampling period of our ADC. The FM-noise spectrum $S_{f_{\text{inst}}}$ is obtained by computing the autocorrelation function of $f_{\text{inst}}(t)$ and by taking its Fourier transform [26], see Fig. 2(c), blue trace. As a reference, we also record the relative phase fluctuations of LO I with respect to LO II, see Fig. 2(c), red trace. The relative phase fluctuations of LO I and LO II are significantly smaller than the phase fluctuations observed in the QD-MLLD measurement, and we can conclude that the linewidth of the LO does not deteriorate our measurements. Fitting the model according to Eq. (1) to the measured FM-noise spectra leads to $S_{\text{L}} = 5.4 \times 10^5$ Hz, $S_1 = 8.4 \times 10^{11}$ Hz$^2$ and $S_2 = 5.0 \times 10^{17}$ Hz$^3$ for the QD-MLLD tone. The measured intrinsic Lorentzian linewidth $\Delta f_{\text{L}}$ is thus 1.9 MHz. We repeat the measurement for different QD-MLLD lines, see Fig. 2(b). To fit Eq. (1), the FM-noise spectrum is smoothed using a moving average with a frequency-dependent window width.

Note that the intrinsic linewidth $\Delta f_{\text{L}}$ can also be inferred from the delay-dependent phase differences $\Delta \phi_\tau(t) = \phi(t+\tau) - \phi(t)$ and the associated phase-noise variance [26]

$$\sigma_\phi^2(\tau) = \left\langle \Delta\phi_\tau(t)^2 \right\rangle. \tag{6}$$

This would involve calculating the slope of $\sigma_\phi^2(\tau)$ close to $\tau = 0$ [26],

$$\Delta f_{\text{L}} = \lim_{\tau \to 0} \frac{\sigma_\phi^2(\tau)}{2\pi\tau}. \tag{7}$$

However, as to be seen in Fig. 2(d), the slope of $\sigma_\phi^2(\tau)$ does not converge to a reliable value for $\tau \to 0$. This scheme is therefore less accurate than the evaluation of power spectrum of the frequency fluctuations according to Eqs. (1)-(4).

### 3.3. Long-term linewidth measurement

The long-term linewidth $\Delta f_{\text{G}}$ is measured using the heterodyne setup shown in Fig. 2(a), Setup I. To this end, we tune LO I close to a comb line of the QD-MLLD and detect the superposition of both on a PD. The resulting photocurrent oscillates at the difference frequencies of LO I and the comb lines. The RF beat note is measured using an ESA having a resolution bandwidth of 200 kHz, which corresponds to an observation time of about $\tau_0 = 15\,\mu\text{s}$. We select different comb lines by tuning LO I, see Fig. 2(b). In these measurements, two important requirements need to be fulfilled: First, the resulting intermediate frequency must be much larger than the expected spectral width of the investigated comb line so that noise-noise interference does not influence the measured linewidth. This is ensured by choosing an intermediate frequency of approximately 1 GHz. Second, the phase fluctuations of local oscillator (LO) must be much smaller than those expected for the investigated comb line. This validity of this assumption was already assured when recording the frequency noise spectrum, see Section 3.2. It is re-confirmed here by measuring the long-term optical linewidth of LO II using Setup I, leading to $\Delta f_G = 0.6$ MHz for the beating of LO I and LO II, see Fig. 2(e), red curve.

For a QD-MLLD tone close to 1547.3 nm, we obtain a long-term linewidth $\Delta f_G$ of approximately 8 MHz, Fig. 2(e), blue curve. The long-term linewidth can also be computed with Eq. (2) using the fitted FM-noise spectra and the observation time $\tau_0 = 15\,\mu\text{s}$, which allows us to estimate $\Delta f_G$ of approximately 7.4 MHz, which fits well with directly measured line shape.

### 4. Coherent transmission using QD-MLLD

In this section, we first investigate the effect of the inherent phase noise of QD-MLLD carriers on coherent data transmission for different modulation formats and symbol rates, see Section 4.1. We show that symbol-wise blind phase search (BPS) algorithm [22], relying on continuous feed-forward estimation and correction of the instantaneous phase difference between the LO tone and the carrier, substantially improves the transmission performance of the QD-MLLD and finally enables the use of 16QAM signaling, see Fig. 3. Details of the phase tracking algorithm are described in Section 4.2. As a reference, we use the transmission performance obtained by block-wise carrier phase recovery (CPR). In this approach, received symbols are processed in blocks, within which a certain frequency offset and an otherwise constant phase difference between the LO tone and the carrier is corrected for [31]. This approach does not allow for dynamic phase tracking and is, e.g., part of the Vector Signal Analyzer (VSA) software offered by Keysight Technologies (89600 series VSA software) [31]. In addition, we benchmark the results by replacing the comb tone by an ECL carrier with an intrinsic linewidth of approximately 10 kHz. We also show that, under realistic transmission conditions, i.e., optical signal-to-noise power ratio (OSNR) levels of 25 dB or less, the carriers of our QD-MLLD comb source yield a transmission performance that is comparable to that of the ECL carrier.

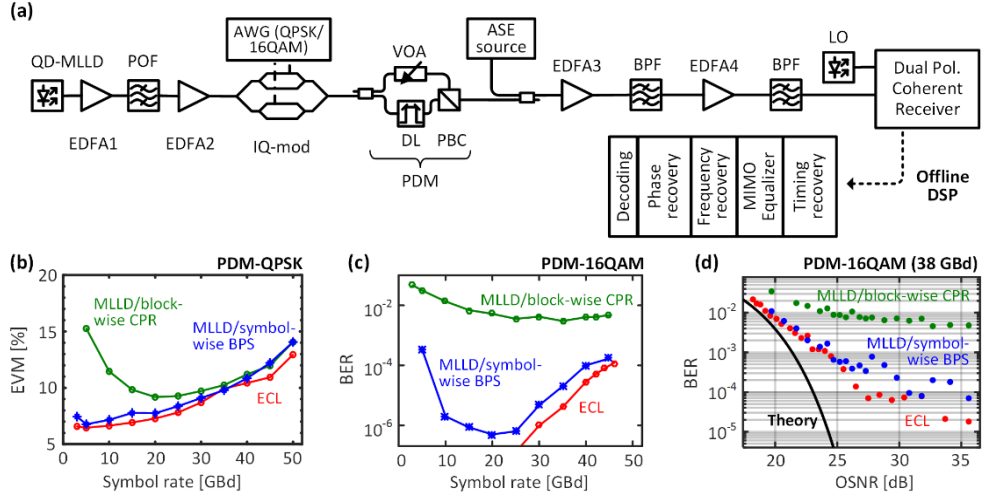

**Fig. 3**. Influence of phase noise on coherent transmission at different symbol rates, modulation formats, and OSNR values. The analysis relies on using a single comb line as a carrier at a wavelength of 1540.4 nm. We compare the performance obtained from the QD-MLLD tone with block-wise carrier phase recovery (MLLD/block-wise CPR) and a symbol-wise blind phase search algorithm (MLLD/symbol-wise BPS). As a reference, we use an optical carrier provided by an ECL with an intrinsic linewidth of ~ 10 kHz. **(a)** Experimental setup. EDFA: Erbium-doped fiber amplifier. POF: Programmable optical filter. IQ-mod: In-phase/quadrature (IQ) modulator. AWG: Arbitrary waveform generator. PDM: Polarization division multiplexing. VOA: Variable optical attenuator. DL: Delay line. PBC: polarization beam combiner. ASE: Amplified spontaneous emission. BPF: Band-pass filter. LO: Local oscillator. **(b)** Error-vector magnitude (EVM) measured for PDM-QPSK signaling at different symbol rates. For QPSK, we did not find sufficient errors within our limited recording lengths and can hence only provide the EVM. We find that symbol-wise BPS algorithm can essentially overcome the phase-noise-related limitations observed for block-wise CPR and bring the transmission performance close to that of a narrowband ECL tone. **(c)** BER results for PDM-16QAM signaling for different symbol rates. For practically relevant symbol rates beyond 30 GBd, symbol-wise BPS can essentially overcome the phase-noise-related transmission performance limitations. **(d)** BER results for PDM-16QAM signaling at 38 GBd for different OSNR values. For OSNR values of 25 dB or less, symbol-wise BPS allows the QD-MLLD tone to perform as good as a narrowband ECL carrier.

In Sections 4.3 and 4.4, we present a series of transmission experiments demonstrating the ability of the symbol-wise BPS scheme to compensate for the phase noise of the QD-MLLD tones. In a first experiment, we transmit 52 channels spaced by 42 GHz over 75 km of standard single-mode fiber (SSMF) using QPSK as a modulation format, see Fig. 4. Using polarization-division-multiplexing (PDM), we achieve an aggregate line rate of 8.32 Tbit/s and a net data rate of 7.83 Tbit/s. In a second experiment, we use 38 carriers to transmit PDM-16QAM signals over 75 km of SSMF at an aggregate line rate of 11.55 Tbit/s and a net data rate of 10.68 Tbit/s, see Fig. 5, which corresponds to the highest data rate achieved by a DC-driven chip-scale comb generator without any additional hardware-based phase-noise reduction [16-19].

*4.1 Influence of comb carrier linewidth and OSNR on coherent communications*

To quantify the influence of the linewidth of the QD-MLLD comb lines on coherent transmission, we measure the BER at different symbol rates for both QPSK and 16QAM. The experimental setup is depicted in Fig. 3(a). After amplification of the frequency comb by EDFA1 to an output power of 17 dBm, we filter the comb line under investigation using a programmable optical filter (POF). A 3dB-bandwidth of ~ 10 GHz of the POF is chosen to effectively suppress all neighboring comb lines. The filtered carrier, centered at a wavelength of 1540.4 nm, is amplified to a constant power of 24 dBm and sent through an IQ modulator. The modulator drive signal is synthesized by an arbitrary-waveform generator (AWG) based on a pseudo-random bit sequence (PRBS) of length $2^{11} - 1$. The bit sequence is mapped to either QPSK or 16QAM symbols with pulses having a raised-cosine (RC) spectrum with a 5% roll-off for QPSK and 10 % for 16QAM. PDM is emulated by splitting the data signal and decorrelating the split signals in time by approximately 5.3 ns with the use of a delay line (DL) [4]. Finally, both signals are combined in orthogonal polarization states using a polarization beam combiner (PBC). To adjust the OSNR of the signal, a noise-loading stage (ASE source) is used. The signal is then sent to the receiver, where it is further amplified by EDFA3 and EDFA4. Bandpass filters (BPF) are used to suppress out-of-band noise after each amplifier. The signal is finally detected by an optical modulation analyzer (OMA, Keysight N4391A) acting as a dual-polarization coherent receiver (DP-CR) with built-in local oscillator (LO) laser. The intrinsic linewidth of the LO is approximately 10 kHz. The output of the DP-CR is digitized by a 4-channel 80 GSa/s real-time oscilloscope (two synchronized Keysight DSO-X 93204A) and recorded for offline digital signal processing (DSP). The digitized signal undergoes a number of DSP steps, comprising timing recovery, MIMO equalization, frequency offset compensation, carrier phase compensation, and decoding, see Section 4.2 for details.

The measured signal quality parameters are summarized in Fig. 3(b) and (c) for QPSK and 16QAM signals. In these plots, green traces represent the results obtained from using block-wise carrier phase recovery ('MLLD/block-wise CPR') as implemented Keysight's VSA software [31]. The blue traces are obtained using continuous feed-forward symbol-wise blind phase search algorithm ('MLLD/symbol-wise BPS'), see Section 4.2 for details. The red traces ('ECL') are obtained by repeating the experiment using a narrowband ECL as an optical source. The ECL is tuned to the same wavelength as the comb tone and has an intrinsic linewidth of approximately 10 kHz, and we can hence assume that the transmission performance is not impaired by phase noise. This is confirmed by observing that block-wise CPR and symbol-wise BPS lead to the same result for the ECL-based transmission. For QPSK, Fig. 3(b), the limited recording length did not contain enough errors for a statistically reliable measurement of the BER. We therefore use the error vector magnitude (EVM) rather than the BER as a performance metric [24]. Assuming that the signals are impaired by additive white Gaussian noise only, the measured EVM would correspond to BER values below $10^{-10}$ [24]. For 16QAM, we measure the BER directly. Generally, the impact of phase noise increases with decreasing symbol rates, i.e., longer optical symbol periods, which lead to an increase of the EVM or the BER at low symbol rates for the QD-MLLD experiments with block-wise CPR, see green traces in Fig. 3(b) and (c). At high symbol rates, the signal quality is impaired by the decreasing OSNR, leading to higher EVM or BER. Using the symbol-wise feed-forward BPS, the impact of phase noise can be greatly reduced, both for QPSK and 16QAM, see blue traces. For QPSK, the transmission performance comes close to that of the narrowband ECL, illustrated by the red trace. 16QAM is generally more sensitive to phase noise than QPSK, since more phases need to be discriminated, and at low symbol rates, even the symbol-wise BPS is not capable of compensating the complete impact of phase noise. Still, the performance of the QD-MLLD experiments with symbol-wise BPS does not show any substantial penalty compared to that of the ECL-based transmission at practically relevant symbol rates of 30 GBd or more.

Finally, we measure the BER as a function of the OSNR, see Fig. 3(d), for 16QAM and a symbol rate of 38 GBd, which is used for the high capacity WDM transmission experiment, see Section 4.4. In this measurement, we again compare the QD-MLLD with block-wise phase recovery (green trace) and symbol-wise phase tracking (blue trace) as well as the ECL reference (red trace). For an OSNR of less than 25 dB and symbol-wise BPS, the penalty of using a QD-MLLD comb line instead of an ECL carrier is below 0.5 dB. Thus, under realistic transmission conditions, the carriers of the QD-MLLD comb source perform similarly as a high-quality individual ECL carrier. The slight impairment arises mainly from residual phase noise that is still left after phase tracking.

*4.2 Data recovery using feed-forward symbol-wise blind phase tracking*

The continuous feed-forward algorithm for symbol-wise phase tracking was implemented in Matlab following the scheme published in [22]. It is part of a multi-step DSP procedure, see Inset of Fig. 3(a). First, the signal is resampled to two samples per symbol followed by timing recovery [32,33]. A 30-tap MIMO equalizer based on the constant-modulus algorithm (CMA) is then used to demultiplex the two orthogonal polarizations of the signal and to compensate for linear transmission impairments [34]. The received symbols from the MIMO equalizer then undergo frequency-offset correction [35]. For QPSK, the $4^{th}$-power of the received symbols is calculated to remove the information of the modulated phase. This leaves only the rapidly increasing phase caused by the frequency offset between signal carrier and LO. The frequency offset is then corrected by fitting a linear curve to the unwrapped phase as a function of time and by subtracting the fitted curve from the measured phase offset. For the case of 16QAM signaling, the $4^{th}$-power scheme can be adapted by QPSK-partitioning [36]. Once the carrier frequency offset is compensated, carrier phase noise compensation is carried out based on a blind phase search (BPS) algorithm [22], which is individually applied to each symbol. The BPS algorithm starts with selecting a group of $N$ complex symbols which are centered around the symbol of interest. The phases of these complex symbols are then modified by adding identical test phases $\phi_t$. The test phases are varied, and the sum of the $N$ vectorial distances to their closest constellation points is minimized with a maximum likelihood (ML) mean square estimation [22]. We apply 45 equidistant test phases for covering an unambiguity range of $\pm 45°$ for an $M$-QAM signal with a quadratic constellation. For a given laser phase noise and a fixed OSNR, an optimum number $N$ of symbols within a group exists, providing an ideal trade-off between strong suppression of phase noise by averaging over many symbols for large $N$ and the ability to track fast phase fluctuations, which requires small $N$. For a symbol rate of 38 GBd, an intrinsic linewidth of the QD-MLLD tone of 1.9 MHz, and OSNR levels of approximately 25 dB, best values of $N$ are typically between 20 and 40. Note that there is a residual probability that phase slips occur in the BPS algorithm. This can be overcome by advanced FEC schemes or pilot tones [37].

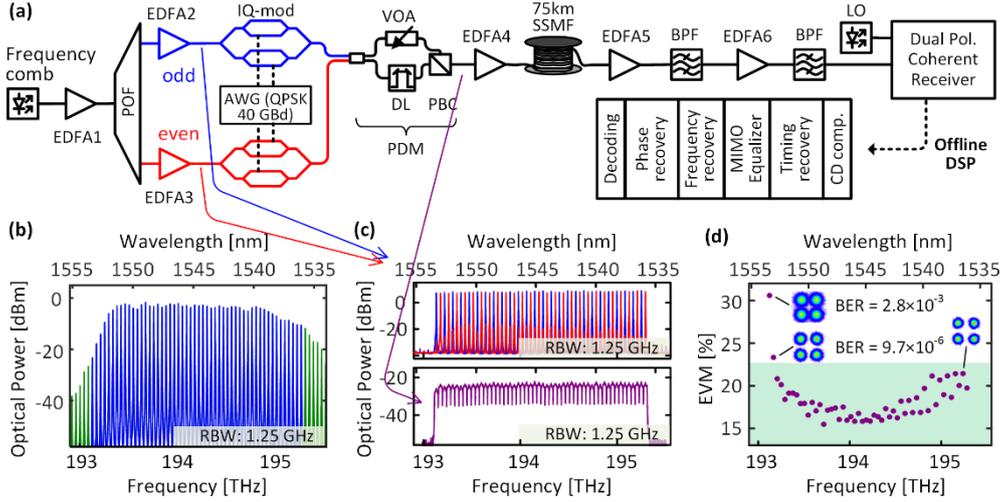

**Fig. 4**. PDM-QPSK data transmission with QD-MLLD. **(a)** Experimental setup used to emulate the WDM transmission experiment. EDFA: Erbium-doped fiber amplifier. POF: Programmable optical filter. AWG: Arbitrary-waveform generator. PDM: Polarization division multiplexing. VOA: Variable optical attenuator. DL: Delay line. PBC: Polarization beam combiner. SSMF: Standard single-mode fiber. BPF: Band-pass filter. LO: Local oscillator. **(b)** Optical spectrum of the QD-MLLD frequency comb. The 52 lines selected for WDM experiment are colored in blue. **(c)** Top: Superimposed spectra of odd and even carriers before modulation. Bottom: Spectrum of 52 modulated carriers. **(d)** EVM of the transmitted WDM channels. The maximum recording length of $2.4 \times 10^{-6}$ bit leads to a minimum detectable BER $> 2.4 \times 10^{-6}$. Within the green shaded area, the BER is smaller.

### 4.3 52× PDM-QPSK 7.83 Tbit/s data transmission

To demonstrate the viability of QD-MLLD in WDM transmission, we perform an experiment at a symbol rate of 40 GBd using QPSK as modulation format. The experimental setup is shown in Fig. 4(a). The total QD-MLLD output power is boosted by EDFA1 to 17 dBm. Using a programmable optical filter (POF), we select 52 tones from the frequency comb, marked in blue in Fig. 4(b), and separate them into odd and even carriers. The two sets of carriers are later encoded with independent data streams to emulate a realistic WDM signal [2,4]. The POF additionally flattens the power spectrum of the carriers. Both carrier sets are amplified by subsequent EDFA2 and EDFA3 and sent through a pair of IQ modulators. Both modulators are driven by QPSK signals with RC spectrum pulses with a 5% roll-off, generated by an AWG using a PRBS of length $2^{11}-1$ at a symbol rate of 40 GBd. After combining the signals, PDM is emulated by a split-and-combine method, described in Section 4.1. The data stream is amplified and transmitted over 75 km of SSMF, see Fig. 4(c) for the power spectrum before and after modulation. At the receiver, we amplify the signal and select each channel using a 0.6 nm tunable BPF followed by EDFA6 and a second tunable 1.5 nm BPF to suppress out-of-band ASE noise. The signal is detected with a DP-CR and undergoes a number of DSP steps, as described in Sections 4.1 and 4.2. To undo the dispersive effect of the 75 km-long fiber link, we additionally perform chromatic dispersion (CD) compensation. The measured EVM values are shown in Fig. 4(d). Due to the limited recording length of $2 \times 10^6$ bit, only the first two channels at the lower spectral edge of the frequency band contain a sufficient number of at least 5 errors to allow for a statistically reliable estimation of the BER. For all other channels, the BER is smaller than the lower measurable limit of $2.4 \times 10^{-6}$ – these data points fall into the green shaded area in Fig. 4(d). Importantly, all 52 channels fall below the BER threshold of $4.7 \times 10^{-3}$ for hard-decision forward-error correction with 6.25% overhead [38]. This leads to an aggregate line rate of 8.32 Tbit/s and a net data rate of 7.83 Tbit/s, transmitted over 75 km of SSMF with a net spectral efficiency of 3.8 bit/s/Hz.

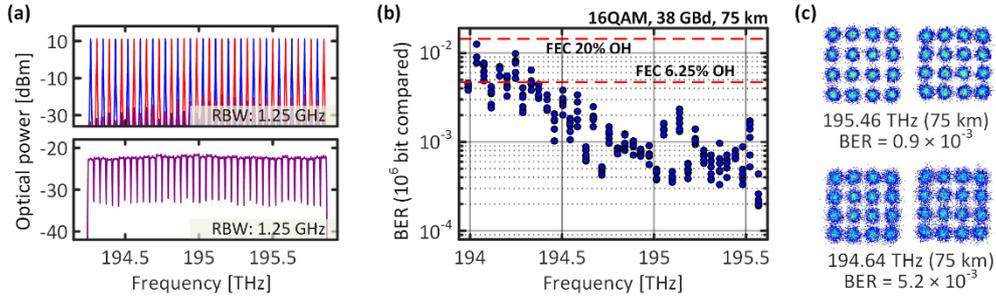

**Fig. 5**. Data transmission results of 38 channels carrying PDM-16QAM signals at 38 GBd. **(a)** Top: Combined odd and even carriers prior to modulation. Bottom: 38 modulated carriers prior to transmission; 304 Gbit/s per channel. **(b)** Measured signal BER in the transmitted channels. **(c)** Exemplary constellation diagrams recorded at different optical carrier frequencies.

### 4.4  38× PDM-16QAM 10.68 Tbit/s data transmission

We also perform a 16QAM transmission demonstration to proof the viability of the symbol-wise blind phase tracking for more complex modulation formats. We use the same setup as for the QPSK experiments, see Fig. 4(a). The symbol rate amounts to 38 GBd, and we use 38 carriers from our QD-MLLD frequency comb for WDM transmission, see Fig. 5(a) for the power spectrum before and after modulation. We use pulses with RC spectrum having a 10% roll-off. We choose the central 38 comb lines, which offer a spectral flatness of better than 3 dB. The measured BER for each of the transmitted WDM channels is depicted in Fig. 5(b). For each channel, five recordings were analyzed, each with $10^6$ bit. Out of the 38 channels, 32 fall below the BER threshold of $4.7 \times 10^{-3}$ for hard-decision forward error correction with 6.25 % overhead, while the remaining six channels fall below the BER threshold of $1.44 \times 10^{-2}$ for forward error correction with 20 % overhead [38]. This leads to an aggregate line rate of 11.55 Tbit/s and a net data rate of 10.68 Tbit/s with a net spectral efficiency of 6.7 bit/s/Hz. To the best of our knowledge, our experiments represent the first demonstration of 16QAM transmission using a DC-driven chip-scale comb generator without any hardware-based phase-noise reduction schemes, and lead to the highest data rate achieved in such an experiment. From Fig. 5(b) one can observe that channels at higher frequencies have a better signal quality. This is due to the fact that our preamplifiers EDFA5 and EDFA6 have a higher noise figure for decreasing optical frequency. Exemplary constellation diagrams for dual polarization channels at carrier frequencies 195.46 THz and 194.64 THz are shown in Fig. 5(c).

### 5.  Summary

We have shown that the strong low-frequency FM noise of QD-MLLD can be overcome by feed-forward symbol-wise phase tracking, thus making the devices usable as particularly simple and attractive multi-wavelength light sources for WDM transmission at data rates beyond 10 Tbit/s. We perform an in-depth analysis of the phase-noise characteristics of QD-MLLD, which affirms a strong increase of the FM-noise spectrum at low frequencies as the main problem that has prevented the use of higher-order modulation formats such as 16QAM so far. To overcome these limitations, we implement and test an advanced digital phase tracking algorithm that is based on symbol-wise blind phase search (BPS) and that finally allows transmission of 38 GBd dual-polarization (DP) 16QAM signals on a total of 38 carriers. This leads to a line rate of 11.55 Tb/s and a net data rate of 10.68 Tbits/s. To the best of our knowledge, this is the highest data rate so far achieved by a DC-driven chip-scale comb generator without any hardware-based phase-noise reduction schemes. We show that under realistic transmission conditions the QD-MLLD tones show a penalty of less than 0.5 dB with respect to a 10 kHz-wide laser carrier.

**Acknowledgements**
This work was supported by the EU-FP7 project BigPipes (# 619591), by the European Research Council (ERC Consolidator Grant 'TeraSHAPE', # 773248), by the Deutsche Forschungsgemeinschaft (DFG) through the Collaborative Research Center 'Wave Phenomena: Analysis and Numerics' (CRC 1173, project B3 'Frequency combs'), by the Karlsruhe School of Optics & Photonics (KSOP), by the Helmholtz Interna-tional Research School for Teratronics (HIRST), and by the Alfried Krupp von Bohlen und Halbach Foun-dation. P.M.-P. was supported by the Erasmus Mundus doctorate programme Europhotonics (grant number 159224-1-2009-1-FR-ERA MUNDUS-EMJD). .